\documentclass[prd,twocolumn,amsmath,showpacs,amssymb,floatfix,nofootinbib,10pt]{revtex4}
\usepackage{indentfirst}
\usepackage{tikz}
\usepackage{setspace}
\usepackage{amsfonts,color,xcolor,graphicx,amsfonts,multirow,amsmath}
\usepackage[section]{placeins}
\definecolor{bluee}{rgb}{0,0,1}
\usepackage[colorlinks=true, colorlinks=true, citecolor=bluee, linkcolor=bluee, urlcolor=bluee]{hyperref}

\allowdisplaybreaks

\begin{document}
\title{Searching for $|V_{cd}|$ through the exclusive decay $D_s^+ \to K^0e^+\nu_e$ within QCD Sum Rules}
\author{Hai-Jiang Tian$^1$}
\author{Yin-Long Yang$^1$}
\author{Dan-Dan Hu$^2$}
\author{Hai-Bing Fu$^1$}
\email{fuhb@gzmu.edu.cn}
\author{Tao Zhong$^1$}
\email{zhongtao1219@sina.com}
\author{Xing-Gang Wu$^2$}
\email{wuxg@cqu.edu.cn}

\address{$^{\it 1}$Department of Physics, Guizhou Minzu University, Guiyang 550025, P.R.China}
\address{$^{\it 2}$Department of Physics, Chongqing Key Laboratory for Strongly Coupled Physics, Chongqing University, Chongqing 401331, P.R.China}

\begin{abstract}
In this paper, we carry out an investigation into the semileptonic decays $D_s^+ \to K^0\ell^+\nu_\ell$ with $\ell=(e,\mu)$ by employing the QCD light-cone sum rules approach. The vector transition form factor (TFF) $f_+^{D_s^+ K^0}(q^2)$ for $D_s^+\to K^0$ decay is calculated while considering its next-to-leading order contribution. Subsequently, we briefly introduce the twist-2, 3 kaon distribution amplitudes, which are calculated by using QCD sum rules within the framework of the background field theory. At the large recoil point, the TFF has $f_+^{D_s^+ K^0}(0)=0.692_{-0.026}^{+0.027}$. Then, we extrapolate $f_+^{D_s^+ K^0}(q^2)$ to the whole physical $q^2$-region via the simplified $z(q^2,t)$-series expansion, and the behavior of TFF $f_+^{D_s^+ K^0}(q^2)$ is exhibited in the numerical results part, including the theoretical and experimental predictions for comparison. In addition, we compute the differential branching fraction $\mathcal{B}(D_s^+ \to K^0\ell^+\nu_\ell)$ with the electron and muon channels, which are expected to be $\mathcal{B}(D_s^+ \to K^0e^+\nu_e)=3.379_{-0.275}^{+0.301}\times 10^{-3}$ and $\mathcal{B}(D_s^+ \to K^0\mu^+\nu_\mu)=3.351_{-0.273}^{+0.299}\times 10^{-3}$ as well as contained other results for comparison. Our results show good agreement with the BESIII measurements and theoretical predictions. Furthermore, we present our prediction with respect to the CKM matrix element $|V_{cd}|$ by using the $\mathcal{B}(D_s^+ \to K^0e^+\nu_e)$ result from BESIII Collaboration, yielding $|V_{cd}|=0.221_{-0.010}^{+0.008}$. Finally, we provide the ratio between $D_s^+ \to K^0e^+\nu_e$ and $D_s^+ \to \eta e^+\nu_e$ channels, \textit{i.e.} $\mathcal{R}_{K^0/\eta}^e=0.144_{-0.020}^{+0.028}$.
\end{abstract}
\date{\today}

\maketitle

\newpage

{\it Introduction.--}Heavy-to-light meson exclusive decay is one of the significant processes in the Standard Model (SM), which plays a crucial role in extracting the Cabibbo-Kobayashi-Maskawa (CKM) matrix elements and in-depth understanding the non-perturbative properties of quantum chromodynamics (QCD). Accurate measurement of the CKM matrix elements has been recognized as a central topic, both theoretically and experimentally. In particular, the Cabibbo-suppressed (semi)-leptonic charm decays are a valuable application for determining the CKM matrix elements $|V_{cd}|$ and $|V_{cs}|$, which have been extensively measured in the BaBar~\cite{BaBar:2007zgf,BaBar:2014xzf}, Belle~\cite{Belle:2006idb}, CLEO~\cite{CLEO:2008ffk,CLEO:2009svp}, and BESIII~\cite{BESIII:2018xre, BESIII:2013iro, BESIII:2015tql, BESIII:2017ylw, BESIII:2018eom, BESIII:2019vhn, Ablikim:2020hsc} Collaborations. The upgraded BESIII has collected an abundance of data samples at the center-of-mass energy of $\sqrt{s}=3.773$, 4.009, 4.13-4.23 and 4.6-4.7 GeV from 2010 to 2023 years, which reported a battery of results with respect to $|V_{cd}|$ and $|V_{cs}|$ by using 2.93 fb$^{-1}$ of data sample adopted at 3.773 GeV. In extracting the CKM matrix elements, which involve the non-perturbative strong effects appearing in the transition from the initial state to the final state, those effects are parameterized by the hadronic invariant form factors. Accordingly, determining the hadronic form factor with greater precision would significantly improve accuracy. In 2018, the BESIII Collaboration reported the first measurements of the form factor for $D_s^+ \to K^0 e^+\nu_e$ and the improved measurements with regard to the absolute branching fraction, which have been made with 3.19~fb$^{-1}$ $e^+e^-$ annihilation data recorded~\cite{BESIII:2018xre}. In addition, they extracted the CKM $|V_{cd}|=0.217\pm0.026\pm0.004$ based on their acquired data sample. The data sample is the biggest that has ever been collected for an experiment in a clean, near-threshold setting for $D_s^+$ investigations.


In the theoretical aspect, for semileptonic decays that involve non-perturbative form factors, one should employ a non-perturbative method to implement the calculations. There are some theoretical predictions concerning semileptonic decays $D_s^+\to K^0\ell^+\nu_\ell$ derived from various techniques, such as the constituent quark model (CQM)~\cite{Melikhov:2000yu}, QCD light-cone sum rules (LCSRs)~\cite{Wu:2006rd}, light-front quark model (LFQM)~\cite{Wang:2008ci,Verma:2011yw}, lattice QCD (LQCD)~\cite{Koponen:2012di}, covariant confined quark model (CCQM)~\cite{Soni:2018adu}, schwinger function methods (SFMs)~\cite{Yao:2021pdy}, and 4-flavor holographic QCD (4FhQCD)~\cite{Ahmed:2023pod}. It is worth noting that, while the application of each individual method is limited, combining them offers a more comprehensive understanding of the highlighted physics.~\cite{Melikhov:2000yu}.

As a well-established theory, the LCSRs method incorporates both the hard and soft contributions in the computation of hadron transitions. This feature enables the LCSRs method to furnish some insights when tackling hadron physics problems spanning various energy scales. Therefore, in this manuscript, we prepare to compute the $D_s^+\to K^0$ transition form factor (TFF) $f_+^{D_s^+ K^0}(q^2)$ based on the QCD LCSRs technique. Additionally, the meson's distribution amplitudes (DAs) are regarded as universal non-perturbative inputs that play a crucial role in the calculations. Therefore, a detailed investigation of them can enhance the theoretical predictions of the QCD LCSRS approach. In our previous works~\cite{Zhong:2022ecl,Zhong:2011rg}, we carefully computed the kaon twist-2 and twist-3 DAs by integrating the phenomenological light-cone harmonic oscillator model with QCD sum rules, all within the framework of background field theory.

{\it Theoretical Framework.--}The $D_s^+\to K^0 \ell^+\nu_\ell$ with $\ell=(e,\mu)$ differential decay width in relation to the mass squared $q^2$ of the $\ell^+\nu_\ell$ system can be written as:
\begin{align}
\frac{d\Gamma(D_s^+\to K^0 \ell^+\nu_\ell)}{dq^2} = \frac{G_F^2|V_{cd}|^2}{24\pi^3}|\vec{p}_K|^3 |f_+^{D_s^+ K^0}(q^2)|^2,
\end{align}
with $G_F=1.166\times 10^{-5}$ GeV$^{-2}$, $\vec{p}_K$ stands for kaon three momentum in the rest frame of $D_s$-meson, and $f_+^{D_s^+ K^0}(q^2)$ refers to the $D_s^+\to K^0$ vector TFF, respectively. To calculate the $D_s^+\to K^0$ TFF's analytical expression within the QCD LCSRs technique, one can start with the following correlation function (also called correlator):
\begin{align}
\Pi _{\mu}(p,q)&=i \int d^4 e^{iq\cdot x}\langle K(p)|T\{j_{\mu}(x), j_{5}^{\dagger}(0) \}| 0 \rangle,
\label{Eq:correlator}
\end{align}
where $j_\mu(x)=\bar{s}\gamma_\mu c(x)$ and $j_5^\dagger(0) =m_c\bar{s}(0)i\gamma_5 c(0)$. The $q$ and $p$ stand for the momentum transfer and kaon momentum. The correlator can be separated into vector and scalar form factors based on the transition matrix element $\langle K^0 (p) |\bar{s}\gamma _{\mu}c| D_s^+ (p+q) \rangle = 2f_+^{D_s^+K^0}(q^2)p_\mu +\tilde{f}^{D_s^+K^0}(q^2)q_\mu$. Meanwhile, the traditional current correlator can well highlight the contributions from the kaon twist-2, 3 DAs, thus better ensuring precision from QCD LCSRs. Since the semileptonic decay amplitudes for $D_s^+\to K^0 \ell^+\nu_\ell$ with $\ell=(e,\mu)$ are solely dependent on the TFF $f_+^{D_s^+ K^0}(q^2)$, we exclusively present the detailed calculation methodology pertaining to this vector TFF. Subsequently, we insert the complete intermediate state, which has the same quantum numbers as the current operator $(\bar{c}i\gamma_5 s)$, into the correlation function in the time-like $q^2$-region and by isolating the pole term of the $D_s^+$-meson ground-state contributions for the invariant amplitude $F_{0(1)}=[q^2,(p+q)^2]$, we can then obtain:
\begin{align}
&f_+^{D_s^+ K^0}(q^2)=\frac{e^{m_{D_s^+}^2/M^2}}{2m_{D_s^+}^2 f_{D_s^+}}
\nonumber\\
&\qquad\times \bigg\{F_0[q^2,M^2,s_0]+\frac{\alpha_s G_F}{4\pi}F_1[q^2,M^2,s_0]\bigg\}.
\end{align}
In which the invariant amplitudes $F_{0(1)}[q^2,M^2,s_0]$ stand for the leading-order (LO) and next-leading-order (NLO) contribution, respectively. Those functions have similar expressions as those of the LO TFF of Hu~\cite{Hu:2021zmy} and the NLO TFF of Duplancic~\cite{Duplancic:2008ix}. As a cross check, we have confirmed that we get the same results via proper transformations. So to short the length of the manuscript, we do not provide the complex expression of $F_{0(1)}[q^2,M^2,s_0]$ here.

As a further step, the kaon twist-2,3 DAs are the key parametric inputs in computation. Since they have already been studied in our past works~\cite{Zhong:2022ecl, Zhong:2011rg}, we shall just briefly describe them here. The basic idea originates from the BHL prescription~\cite{Huang:1994dy}, which can link the equal-time wavefunction (WF) in the rest frame and the light-cone WF and then use the acquired WF to construct the twist-2, 3 DAs. The relationship between the kaon twist-2, 3 DAs and their WFs can be written as $\phi_{i;K^0}(x,\mu) = {\cal N}_i \int_{|\mathbf{k}_\bot| \leqslant \mu^2} \frac{d^2\mathbf{k}_\bot}{16\pi^3} \Psi_{i;K^0}(x,\mathbf{k}_\bot)$, where the index $i = (2,3)$ stand for twist-2, 3 DAs respectively. The normalization coefficients are ${\cal N}_2 = 2\sqrt{6}/f_K$ and ${\cal N}_3 = 1$. Then, the twist-2, 3 WFs are expressed as,
\begin{align}\label{WF_twist-2} \nonumber
&\Psi_{2;K^0}(x,\mathbf{k}_\bot)=\frac{\tilde{m}}{\sqrt{\mathbf{k}_\bot^2 +\tilde{m}^2}} A_{2;K^0}\varphi_{2;K^0}(x) \\
&\quad \times\exp{\left[ -\frac{1}{8\beta_{2:K^0}^2}\left( \frac{\mathbf{k}_\bot^2+\hat{m}_s^2}{x}+ \frac{\mathbf{k}_\bot^2+\hat{m}_q^2}{\bar{x}}\right)\right]},
\end{align}
with $\varphi _{2;K^0}(x) =\sum_{n=1}^2{[x\bar{x}]^{\alpha_{2;K^0}}[1+\hat{B}^n_{2;K^0} C_{n}^{3/2}(\xi)]}$ dominates the WF's longitudinal distribution, and
\begin{align}\label{WF_twist-3} \nonumber
&\Psi _{3;K^0}^{(p,\sigma)}( x,\mathbf{k}_{\bot}) =\frac{A_{3;K^0}^{(p,\sigma)}\varphi _{3;K^0}^{(p,\sigma)}(x)}{x\bar{x}}  \\
&\quad \times\exp \left[ -\frac{1}{8(\beta _{3;K^0}^{(p,\sigma)}) ^2}\left( \frac{\hat{m}_{q}^{2}+\mathbf{k}_{\bot}^{2}}{x}+\frac{\hat{m}_{s}^{2}+\mathbf{k}_{\bot}^{2}}{\bar{x}} \right) \right],
\end{align}
with $\varphi_{3;K^0}^{(p,\sigma)}(x) =\sum_{n=1}^2{\left[ 1+\hat{B}_{3;K^0}^{(p,\sigma),n}C_{n}^{(1/2 ,3/2)}(\xi)\right]}$. The tags that appear above have the following descriptions: $\bar{x}=(1-x)$, $\xi = (2x-1)$, $\tilde{m}=(\hat{m}_q x+ \hat{m}_s\bar{x})$ with $\hat{m}_q$ and $\hat{m}_s$ are the constituent quark masses, $\mathbf{k}_\bot$ stands for the transverse momentum, $C_n^{(1/2,3/2)}(\xi)$ represent for the Gegenbauer polynomial, while other marks refer to the WF's model parameters. Note that we set $\hat{B}^1_{2;K^0}=0.4\hat{B}^2_{2;K^0}$ where the factor 0.4 comes from the ratio of the first and second Gegenbauer moments. Those WF's model parameters can be fitted through the kaon twist-2, 3 DAs $\langle\xi^n\rangle$-moments, which are computed by using QCD sum rules within the framework of background field theory. Usually, we start with the related correction function to achieve the final formula undergoing the finest calculation. Furthermore, for the twist-4 DAs, we chose them in Ref.~\cite{Ball:2006wn}.

{\it Numerical Analysis.--}In processing the numerical analysis, we adopt the following basic input parameters from Particle Data Group (PDG)~\cite{ParticleDataGroup:2022pth}: $m_{D_s^+} =1968.35\pm0.07$ MeV, $m_{K^0}=497.611\pm0.013$ MeV, the $d$ and $s$-quark masses are $m_d=4.67_{-0.17}^{+0.48}$ MeV and $m_s=93_{-3.4}^{+8.6}$ MeV, while the charm quark mass is $\hat{m}_c=1.27\pm0.02$ GeV. The $D_s^+$ and $K$ meson decay constants are taken as $f_{D_s^+}=251.1\pm2.4\pm3.0$ MeV~\cite{BESIII:2021bdp} and $f_K/f_\pi=1.1932$~\cite{FlavourLatticeAveragingGroup:2019iem} with $f_\pi=130.2(1.2)$ MeV~\cite{ParticleDataGroup:2022pth}.

In this part, we directly exhibit the numerical values of the twist-2, 3 DAs $\left<\xi^n \right>$-moments in Table~\ref{Tab:Xin}, whose specific calculation process can be seen in Refs.~\cite{Zhong:2022ecl, Zhong:2011rg}. By means of those numerical results, one can fit the model parameters of the Eqs.~(\ref{WF_twist-2}) and (\ref{WF_twist-3}) so as to proceed with the next calculation. Subsequently, we give the model parameters's results corresponding to the applicable energy scale $(m_{D_s}^2 -m_c^2)^{1/2}\approx 1.5$ GeV for TFF $f_+^{D_s^+ K^0}(q^2)$,
\begin{align}
    &A_{2;K^0}=15.93,&&\alpha_{2;K^0}=0.402,\nonumber\\
    &\hat{B}^2_{2;K^0}=0.007,&&\beta_{2;K^0}=2.660,
\end{align}
for twist-2 kaon DA. Meanwhile, the parameters for kaon twist-3 DAs are
\begin{align}
    &A_{3;K^0}^p=133.6,&&\hat{B}_{3;K^0}^{p,1}=-0.002,\nonumber\\
    &\hat{B}_{3;K^0}^{p,2}=1.602,&&\beta_{3;K^0}^p=0.547,\nonumber\\
    &A_{3;K^0}^\sigma=153.1,&&\hat{B}_{3;K^0}^{\sigma,1}=0.015,\nonumber\\
    &\hat{B}_{3;K^0}^{\sigma,2}=0.223,&&\beta_{3;K^0}^\sigma=0.471.
\end{align}

\begin{table}[t]
\footnotesize
\begin{center}
\caption{Results of the kaon twist-2,3 DAs $\langle\xi^n\rangle$-moments at the scale $\mu_0=1$ with $n=(1,...,10)$ for twist-2 and $n=(1,2)$ for twist-3.}
\label{Tab:Xin}
\begin{tabular}{ll}
\hline
\hline
$\hspace{1.6cm}\mathrm{Odd}$~~~~~~~~~~~~~~~~~~~~~~~~~~&$\hspace{1.6cm}\mathrm{Even}$          \\
\hline
$\langle\xi^1\rangle_{2;K^0}|_{\mu_0}=-0.0438_{-0.0075}^{+0.0053}$                  &$\langle\xi^2\rangle_{2;K^0}|_{\mu_0}=+0.262_{-0.010}^{+0.010}$                    \\
$\langle\xi^3\rangle_{2;K^0}|_{\mu_0}=-0.0210_{-0.0035}^{+0.0024}$                  &$\langle\xi^4\rangle_{2;K^0}|_{\mu_0}=+0.132_{-0.006}^{+0.006}$                     \\
$\langle\xi^5\rangle_{2;K^0}|_{\mu_0}=-0.0134_{-0.0021}^{+0.0014}$                  &$\langle\xi^6\rangle_{2;K^0}|_{\mu_0}=+0.082_{-0.005}^{+0.005}$                      \\
$\langle\xi^7\rangle_{2;K^0}|_{\mu_0}=-0.0087_{-0.0014}^{+0.0009}$                  &$\langle\xi^8\rangle_{2;K^0}|_{\mu_0}=+0.058_{-0.004}^{+0.004}$                       \\
$\langle\xi^9\rangle_{2;K^0}|_{\mu_0}=-0.0058_{-0.0010}^{+0.0007}$                  &$\langle\xi^{10}\rangle_{2;K^0}|_{\mu_0}=+0.044_{-0.004}^{+0.004}$                        \\
\hline
$\langle\xi_p^1\rangle_{3;K^0}|_{\mu_0}=-0.126_{-0.050}^{+0.034}$                    &$\langle\xi_p^2\rangle_{3;K^0}|_{\mu_0}=+0.425_{-0.064}^{+0.063}$                          \\
$\langle\xi_\sigma^1\rangle_{3;K^0}|_{\mu_0}=-0.094_{-0.044}^{+0.030}$               &$\langle\xi_\sigma^2\rangle_{3;K^0}|_{\mu_0}=+0.329_{-0.052}^{+0.057}$                      \\
\hline
\hline
\end{tabular}
\end{center}
\end{table}

\begin{table}[b]
\footnotesize
\begin{center}
\caption{Comparison of $f_{+}^{D_s^+ K^0}(0)$ for $D_s^+\to K^0$ transition.}
\label{Tab:TFFs}
\begin{tabular}{l l}
\hline
\hline
~~~~~~~~~~~~~~~~~~~~~~~~~~~~~~~~~~~~~~~~~~~~~~~~~~~~~~~~&$f_{+}^{D_s^+ K^0}(0)$           \\
\hline
This Work                                      &$0.692_{-0.026}^{+0.027}$                                               \\
BESIII'18-I~\cite{BESIII:2018xre}              &$0.720\pm0.084\pm0.013$                                                  \\
CQM'00~\cite{Melikhov:2000yu}                  &$0.72\pm0.20$                                                             \\
LCSR'06~\cite{Wu:2006rd}                       &$0.820_{-0.071}^{+0.080}$                                                  \\
LFQM'08~\cite{Wang:2008ci}                     &$0.67$                                                                      \\
LFQM'12~\cite{Verma:2011yw}                    &$0.66$                                                                       \\
CCQM'18~\cite{Soni:2018adu}                    &$0.60\pm0.09$                                                                 \\
SFMs'21~\cite{Yao:2021pdy}                     &$0.681(10)$                                                                    \\
4FhQCD'23~\cite{Ahmed:2023pod}                 &$0.57$                                                                          \\
\hline
\hline
\end{tabular}
\end{center}
\end{table}

Furthermore, to obtain the numerical results for $D_s^+\to K^0$ TFF, one need to determine two parameters, $\it i.e.,$ the continuum threshold $s_0$ and Borel parameters $M^2$. We can determine them based on the basic criteria of the sum rules approach~\cite{Tian:2023vbh}: the continuum threshold $s_0=6.7(0.5)$ GeV$^2$ is taken to be close to the squared mass of the $D_s^+$-meson excited state, $D_{s0}(2590)$, and the suitable Borel window is found to be $M^2=14(1)$. According to those determined parameters, we present the $D_s^+\to K^0$ TFF at the large recoil point $f_+^{D_s^+ K^0}(0)$, including other results for comparison, $\it e.g.,$ the BESIII'18-I~\cite{BESIII:2018xre}, CQM'00~\cite{Melikhov:2000yu}, LCSR'06~\cite{Wu:2006rd}, LFQM'08(12)~\cite{Wang:2008ci,Verma:2011yw}, CCQM'18~\cite{Soni:2018adu}, SFMs'21~\cite{Yao:2021pdy}, and 4FhQCD'23~\cite{Ahmed:2023pod}, which are listed in Table~\ref{Tab:TFFs}. Upon examining the Table~\ref{Tab:TFFs}, it is evident that the predictions fall within the range of $0.5 \sim 0.9$, and our results lie in the upper-middle section, demonstrating a good degree of agreement. In addition, the contribution proportion of the NLO is approximately $12\%$ in our total results.

\begin{figure}[t]
\begin{center}
\includegraphics[width=0.425\textwidth]{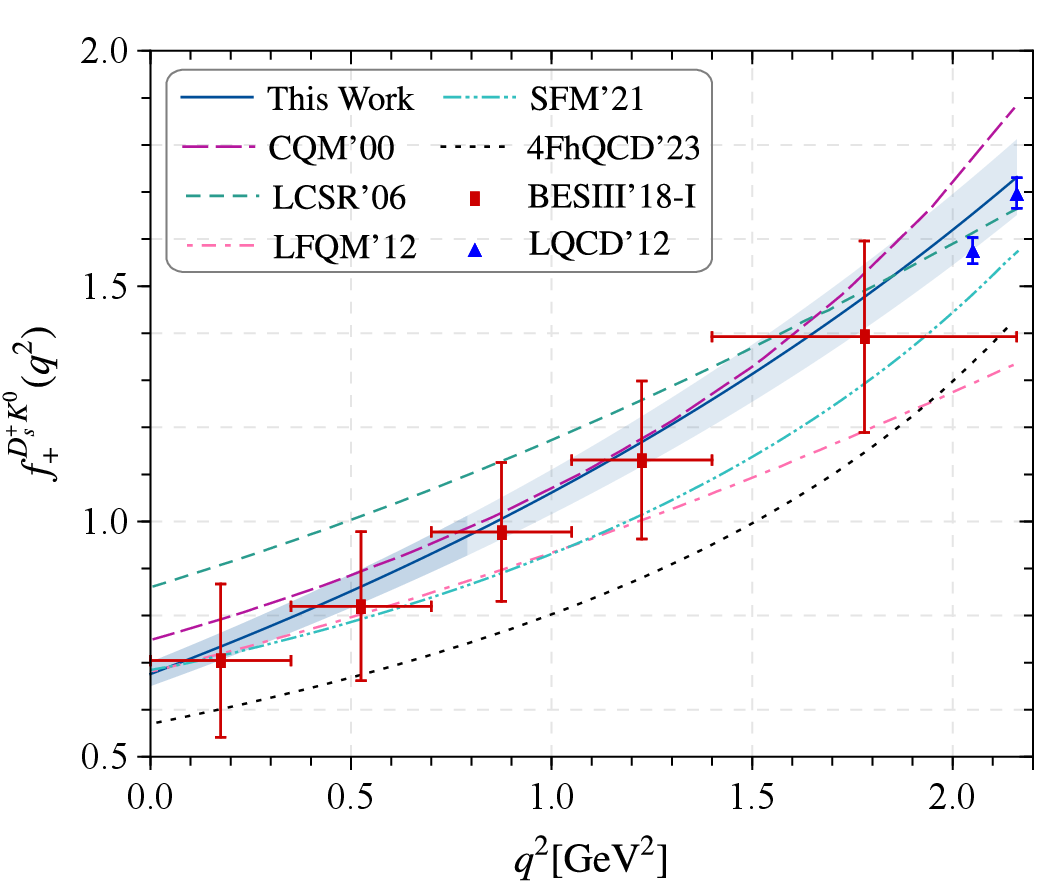}
\end{center}
\caption{The curves of $D_s^+\to K^0$ TFF $f_+^{D_s^+ K^0}(q^2)$ along with associated uncertainties, is done throughout the whole $q^2$-region. Compared with various results from the theoretical and experimental, $\it i.e.,$ the BESIII'18-I~\cite{BESIII:2018xre}, CQM'00~\cite{Melikhov:2000yu}, LCSR'06~\cite{Wu:2006rd}, LFQM'12~\cite{Verma:2011yw}, LQCD'12~\cite{Koponen:2012di}, SFMs'21\cite{Yao:2021pdy}, and 4FhQCD'23~\cite{Ahmed:2023pod}. The results of 4FhQCD'23 cited here were obtained from $f_+^{D_s^+ K^0}(q^2)/f_+^{D_s^+ K^0}(0)$ multiplying $f_+^{D_s^+ K^0}(0)$~\cite{Ahmed:2023pod}.}
\label{Fig:TFFs_fp}
\end{figure}
\begin{figure}[t]
\begin{center}
\includegraphics[width=0.425\textwidth]{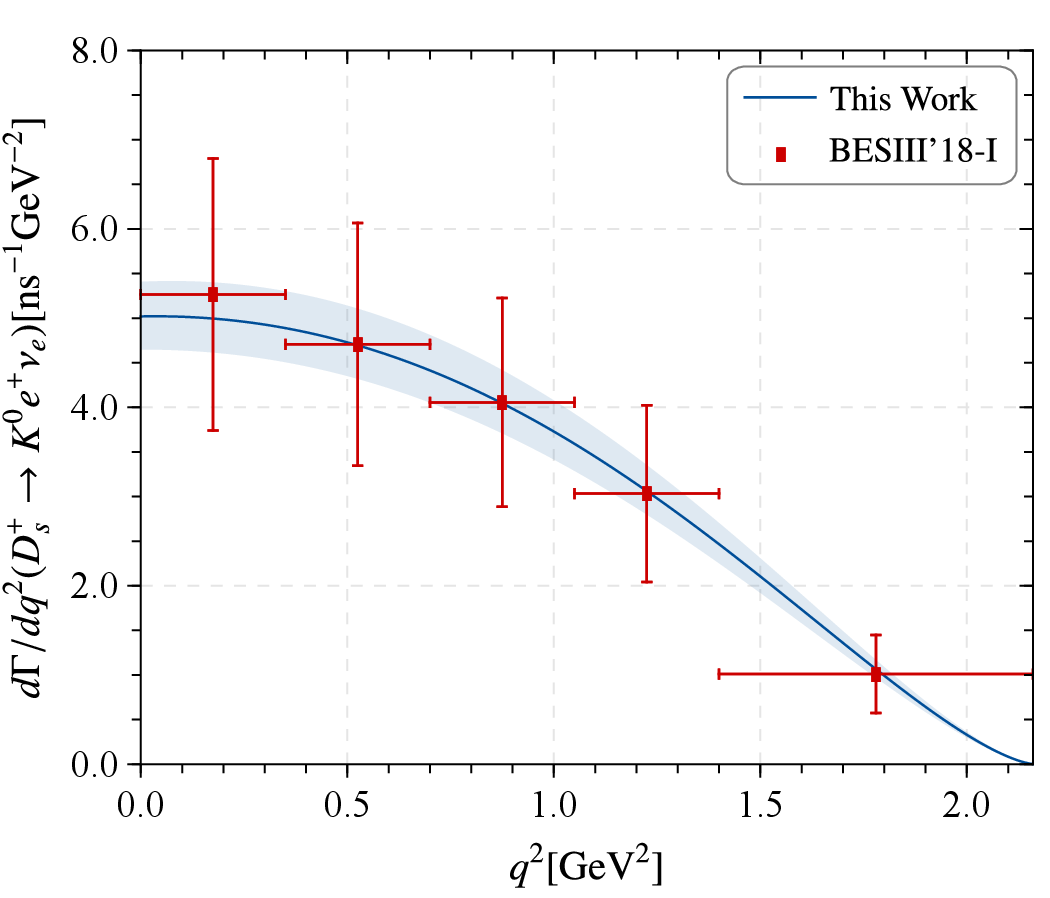}
\end{center}
\caption{Differential decay rates for $D_s^+\to K^0 e^+\nu_e$ as a function of the electron invariant mass squared($q^2$) with the shaded bands stand for its uncertainties, including the experimental results for comparison from the BESIII'18-I~\cite{BESIII:2018xre}.}
\label{Fig:dB}
\end{figure}

The physical allowable region for the $D_s^+\to K^0$ TFF is approximately $q^2 \leqslant (m_{D_s^+}-m_{K^0})^2 \approx 2.16$ GeV$^2$. Then, according to the applicable $q^2$-region of the LCSRs approach, $\it i.e.,$ the low and intermediate region, we take $q^2 \in [0,0.8]$ GeV$^2$, while the high $q^2$-region can be extrapolated through $z(q^2,t)$ converging the simplified series expansion (SSE). The $D_s^+\to K^0$ TFF can expand as:
\begin{eqnarray}
f_+^{D_s^+ K^0}\left( q^2 \right) =\frac{1}{1-q^2/m_{H}^{2}}\sum_{k=0,1,2}{\beta _kz^k\left( q^2,t \right)},
\end{eqnarray}
in which $H$ is $D_s^+$-meson, $\beta_k$ refers to the fitting parameters, while the function $z(q^2,t)=(\sqrt{t_+-q^2}-\sqrt{t_+-t_0})/(\sqrt{t_+-q^2}+\sqrt{t_+-t_0})$ with $t_\pm=(m_{D_s^+}\pm m_{K^0})^2$ and $t_0=t_\pm(1-\sqrt{1-t_-/t_+})$. The SSE method can remain the analytic structure correct in the complex plane and assure the suitable scaling, $f_+^{D_s^+ K^0}(q^2)\sim 1/q^2$.

After extrapolating the $D_s^+\to K^0$ TFF to the complete physical $q^2$-region, one can obtain the entire behavior of $f_+^{D_s^+ K^0}(q^2)$, which is exhibited in Fig.~\ref{Fig:TFFs_fp}. In which, the solid line is the central behavior of $f_+^{D_s^+ K^0}(q^2)$, the darker band stands for the LCSRs prediction, and the lighter band refers to the SSE result. Meanwhile, we also present the theoretical and experimental results for comparison from the CQM'00~\cite{Melikhov:2000yu}, LCSR'06~\cite{Wu:2006rd}, LFQM'12~\cite{Verma:2011yw}, LQCD'12~\cite{Koponen:2012di}, BESIII'18-I~\cite{BESIII:2018xre}, SFMs'21~\cite{Yao:2021pdy}, and 4FhQCD'23~\cite{Ahmed:2023pod}.

By means of the determined $D_s^+\to K^0$ TFF from the LCSRs, one can further proceed with the calculation with respect to the physical observables of the semi-leptonic decay $D_s^+\to K^0 \ell^+\nu_\ell$ with $\ell=(e,\mu)$. Then, we presented the integrated values of differential decay widths for $D_s^+\to K^0 \ell^+\nu_\ell$ with two channels:
\begin{eqnarray}
&\mathrm{\Gamma}(D_s^+\to K^0e^+\nu_e)=4.413_{-0.359}^{+0.393}\times10^{-15}~\mathrm{GeV},\\
&\mathrm{\Gamma}(D_s^+\to K^0\mu^+\nu_\mu)=4.376_{-0.357}^{+0.390}\times10^{-15}~\mathrm{GeV}.
\end{eqnarray}
As a further step, we can obtain the branching fraction results through the utilization of the lifetime of the initial state $D_s^+$-meson, $\it i.e.,$ $\tau_{D_s^+}=0.504\pm0.004$ ps~\cite{ParticleDataGroup:2022pth}. The predictions are listed in Table~\ref{Tab:BR}, including the theoretical and experimental results for a comparison, $\it e.g.,$ the BESIII'18-I~\cite{BESIII:2018xre}, LCSR'06~\cite{Wu:2006rd}, LFQM'08~\cite{Wang:2008ci}, CCQM'18~\cite{Soni:2018adu}, PDG'22~\cite{ParticleDataGroup:2022pth}, HIETALA'15~\cite{Hietala:2015jqa}, LFQM'17~\cite{Cheng:2017pcq}, and CLEO'09-I~\cite{CLEO:2009dyb}. In which, our numerical results can match with those experimental results within their uncertainties, however, the statistical uncertainties remain significantly large, and thus it is necessary to update the more precise experimental results in the near future. In Fig.~\ref{Fig:dB}, we exhibit the behavior for the differential width of $D_s^+\to K^0e^+\nu_e$ in whole $q^2$-region, including the experimental results for comparison.

\begin{table}[t]
\footnotesize
\begin{center}
\caption{The branching fractions (in unit: $10^{-3}$) for $D_s^+\to K^0e^+\nu_e$ and $D_s^+\to K^0\mu^+\nu_\mu$ channels. For comparison, the theoretical and experimental results are also presented. }
\label{Tab:BR}
\begin{tabular}{l l l}
\hline
\hline
~~~~~~~~~~~~~~~~~~~~~~~~~~&$\mathcal{B}(D_s^+\to K^0e^+\nu_e)$~~~~~~~&$\mathcal{B}(D_s^+\to K^0\mu^+\nu_\mu)$           \\
\hline
This Work                                        &$3.379_{-0.275}^{+0.301}$                          &$3.351_{-0.273}^{+0.299}$                          \\
BESIII'18-I~\cite{BESIII:2018xre}                &$3.25\pm0.38\pm0.16$                            &-                                                \\
LCSR'06~\cite{Wu:2006rd}                         &$3.90_{-0.57}^{+0.74}$                          &$3.83_{-0.56}^{+0.72}$                            \\
LFQM'08~\cite{Wang:2008ci}                       &$2.9$                                           &-                                                  \\
CCQM'18~\cite{Soni:2018adu}                      &$2$                                             &$2$                                                 \\
PDG'22~\cite{ParticleDataGroup:2022pth}          &$0.34\pm0.04$                                                                                         \\
HIETALA'15~\cite{Hietala:2015jqa}                &$3.9\pm0.8\pm0.3$                               &-                                                     \\
LFQM'17~\cite{Cheng:2017pcq}                     &$0.27\pm0.02$                                   &$0.26\pm0.02$                                          \\
CLEO'09-I~\cite{CLEO:2009dyb}                    &$3.7\pm1\pm0.2$                                 &-                                                       \\
\hline
\hline
\end{tabular}
\end{center}
\end{table}
\begin{figure}[b]
\begin{center}
\includegraphics[width=0.425\textwidth]{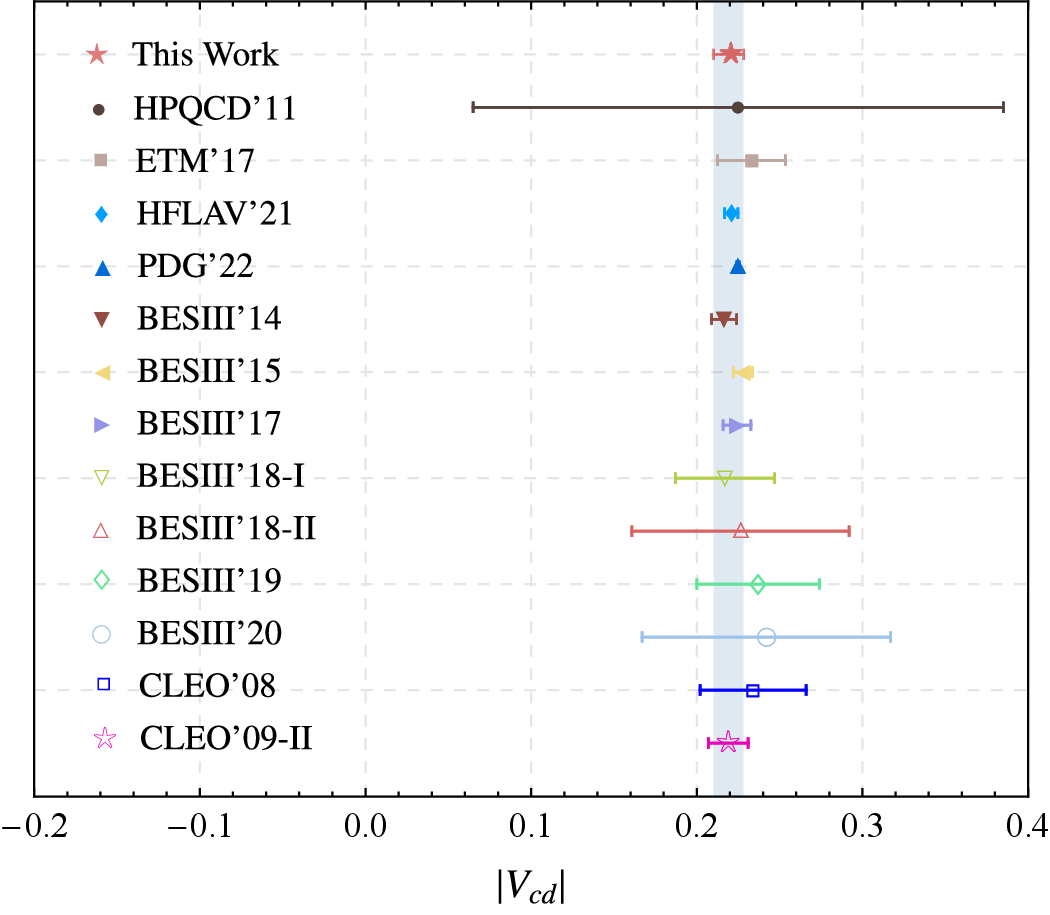}
\end{center}
\caption{Our prediction of $|V_{cd}|$ obtained from the decay channel $D_s^+ \to K^0 e^+\nu_e$, compared with other theoretical and experimental results of various channels.}
\label{Fig:Vcd}
\end{figure}

Furthermore, to provide a new result on $|V_{cd}|$ within the QCD LCSRs approach, we extract the CKM matrix element $|V_{cd}|$ based on the branching fractions $\mathcal{B}(D_s^+\to K^0e^+\nu_e)$ from the BESIII'18-I~\cite{BESIII:2018xre}, which is listed as follows:
\begin{eqnarray}
|V_{cd}|=0.221_{-0.010}^{+0.008}.
\end{eqnarray}
In addition, for comparison, we gathered the CKM matrix element $|V_{cd}|$ results from the theoretical and experimental studies, which are presented in Fig.~\ref{Fig:Vcd}. Those results originate from the CLEO'08(09-II)~\cite{CLEO:2008ffk,CLEO:2009svp}, BESIII'(14-20)~\cite{BESIII:2018xre,BESIII:2013iro,BESIII:2015tql,BESIII:2017ylw,BESIII:2018eom,BESIII:2019vhn,Ablikim:2020hsc}, PDG'22~\cite{ParticleDataGroup:2022pth}, HPQCD'11~\cite{Na:2011mc}, ETM'17~\cite{Lubicz:2017syv}, and HFLAV'21~\cite{HFLAV:2022esi}. And our theoretical prediction aligns well with the experimental outcomes.

Lastly, we attempt to give a theoretical ratio $\mathcal{R}_{K^0/\eta}^e$ with respect to the different decay channels as follows:
\begin{align}
    \mathcal{R}_{K^0/\eta}^e &= \frac{\mathcal{B}(D_s^+\to K^0 e^+\nu_e)}{\mathcal{B}(D_s^+\to \eta e^+\nu_e)} \nonumber\\
    &=0.144_{-0.020}^{+0.028}.
\end{align}
The result $\mathcal{B}(D_s^+\to \eta e^+\nu_e)=2.346_{-0.331}^{+0.418}\times 10^{-2}$ comes from our previous work~\cite{Hu:2021zmy}. The ratio presented can be viewed as a test for the $D_s^+$-meson's internal structure and the properties of related pseudoscalar mesons.

{\it Summary.--}In this paper, we conducted an investigation into the semileptonic decays $D_s^+\to K^0\ell^+\nu_\ell$ with $\ell=(e,\mu)$. Firstly, we calculated the $D_s^+\to K^0$ vector TFF $f_+^{D_s^+ K^0}(q^2)$ by using the QCD LCSRs approach up to the NLO correction and briefly introduced the kaon twist-2, 3 DAs. Then we plotted the behavior of the $D_s^+\to K^0$ TFF $f_+^{D_s^+ K^0}(q^2)$ in the whole $q^2$-region, which is exhibited in Fig.~\ref{Fig:TFFs_fp}, including the theoretical and experimental results for comparison. Meanwhile, according to the derived TFF $f_+^{D_s^+ K^0}(q^2)$, we also computed the differential width of $D_s^+\to K^0e^+\nu_e$ and presented it in Fig.~\ref{Fig:dB}. As the further step, we calculated the branching fractions of the semileptonic decays $D_s^+\to K^0\ell^+\nu_\ell$ with $\ell=(e,\mu)$ by means of the lifetime of the initial state $D_s^+$-meson, which are $\mathcal{B}(D_s^+ \to K^0e^+\nu_e)=3.379_{-0.275}^{+0.301}\times 10^{-3}$ and $\mathcal{B}(D_s^+ \to K^0\mu^+\nu_\mu)=3.351_{-0.273}^{+0.299}\times 10^{-3}$. Furthermore, we extracted the CKM matrix element $|V_{cd}|$ by using the $\mathcal{B}(D_s^+ \to K^0e^+\nu_e)$ from the BESIII Collaboration, which is expected to be $|V_{cd}|=0.221_{-0.010}^{+0.008}$. Lastly, we presented a theoretical ratio $\mathcal{R}_{K^0/\eta}^e=0.144_{-0.020}^{+0.028}$.
\\
\\
{\it Acknowledgments.--}This work was supported in part by the National Natural Science Foundation of China under Grant No.12265010, No.12265009, No.12175025 and No.12347101, the Project of Guizhou Provincial Department of Science and Technology under Grant No.ZK[2021]024 and No.ZK[2023]142.


\begin{thebibliography}{99}


\bibitem{BaBar:2007zgf}
    B.~Aubert \textit{et al.} [BaBar Collaboration],
    \href{https://doi.org/10.1103/PhysRevD.76.052005}
    {Phys. Rev. D \textbf{76}, 052005 (2007)}.



\bibitem{BaBar:2014xzf}
    J.~P.~Lees \textit{et al.} [BaBar Collaboration],
    \href{https://doi.org/10.1103/PhysRevD.91.052022}
    {Phys. Rev. D \textbf{91}, no.5, 052022 (2015)}.



\bibitem{Belle:2006idb}
    L.~Widhalm \textit{et al.} [Belle Collaboration],
    \href{https://doi.org/10.1103/PhysRevLett.97.061804}
    {Phys. Rev. Lett. \textbf{97}, 061804 (2006)}.




\bibitem{CLEO:2008ffk}
    B.~I.~Eisenstein \textit{et al.} [CLEO Collaboration],
    \href{https://doi.org/10.1103/PhysRevD.78.052003}
    {Phys. Rev. D \textbf{78}, 052003 (2008)}.




\bibitem{CLEO:2009svp}
    D.~Besson \textit{et al.} [CLEO Collaboration],
    \href{https://doi.org/10.1103/PhysRevD.80.032005}
    {Phys. Rev. D \textbf{80}, 032005 (2009)}.



\bibitem{BESIII:2015tql}
    M.~Ablikim \textit{et al.} [BESIII Collaboration],
    \href{https://doi.org/10.1103/PhysRevD.92.072012}
    {Phys. Rev. D \textbf{92}, no.7, 072012 (2015)}.


\bibitem{Ablikim:2020hsc}
    M.~Ablikim [BESIII Collaboration],
    \href{https://doi.org/10.1103/PhysRevLett.124.231801}
    {Phys. Rev. Lett. \textbf{124}, no.23, 231801 (2020)}.


\bibitem{BESIII:2013iro}
    M.~Ablikim \textit{et al.} [BESIII Collaboration],
    \href{https://doi.org/10.1103/PhysRevD.89.051104}
    {Phys. Rev. D \textbf{89}, no.5, 051104 (2014)}.


\bibitem{BESIII:2017ylw}
    M.~Ablikim \textit{et al.} [BESIII Collaboration],
    \href{https://doi.org/10.1103/PhysRevD.96.012002}
    {Phys. Rev. D \textbf{96}, no.1, 012002 (2017)}.


\bibitem{BESIII:2018eom}
    M.~Ablikim \textit{et al.} [BESIII Collaboration],
    \href{https://doi.org/10.1103/PhysRevD.97.092009}
    {Phys. Rev. D \textbf{97}, no.9, 092009 (2018)}.


\bibitem{BESIII:2018xre}
    M.~Ablikim \textit{et al.} [BESIII Collaboration],
    \href{https://doi.org/10.1103/PhysRevLett.122.061801}
    {Phys. Rev. Lett. \textbf{122}, 061801 (2019)}.

\bibitem{BESIII:2019vhn}
    M.~Ablikim \textit{et al.} [BESIII Collaboration],
    \href{https://doi.org/10.1103/PhysRevLett.123.211802}
    {Phys. Rev. Lett. \textbf{123}, no.21, 211802 (2019)}.

\bibitem{Melikhov:2000yu}
    D.~Melikhov and B.~Stech,
    \href{https://doi.org/10.1103/PhysRevD.62.014006}
    {Phys. Rev. D \textbf{62}, 014006 (2000)}.

\bibitem{Wu:2006rd}
    Y.~L.~Wu, M.~Zhong and Y.~B.~Zuo,
    \href{https://doi.org/10.1142/S0217751X06033209}
    {Int. J. Mod. Phys. A \textbf{21}, 6125-6172 (2006)}.


\bibitem{Wang:2008ci}
    W.~Wang and Y.~L.~Shen,
    \href{https://doi.org/10.1103/PhysRevD.78.054002}
    {Phys. Rev. D \textbf{78}, 054002 (2008)}.


\bibitem{Verma:2011yw}
    R.~C.~Verma,
    \href{https://doi.org/10.1088/0954-3899/39/2/025005}
    {J. Phys. G \textbf{39}, 025005 (2012)}.


\bibitem{Koponen:2012di}
    J.~Koponen \textit{et al.} [HPQCD Collaboration],
    \href{https://arxiv.org/abs/1208.6242}
    {arXiv:1208.6242}.


\bibitem{Soni:2018adu}
    N.~R.~Soni, M.~A.~Ivanov, J.~G.~K\"orner, J.~N.~Pandya, P.~Santorelli and C.~T.~Tran,
    \href{https://doi.org/10.1103/PhysRevD.98.114031}
    {Phys. Rev. D \textbf{98}, 114031 (2018)}.


\bibitem{Yao:2021pdy}
    Z.~Q.~Yao, D.~Binosi, Z.~F.~Cui and C.~D.~Roberts,
    \href{https://doi.org/10.1016/j.physletb.2021.136793}
    {Phys. Lett. B \textbf{824}, 136793 (2022)}.

\bibitem{Ahmed:2023pod}
    H.~A.~Ahmed, Y.~Chen and M.~Huang,
    \href{https://doi.org/10.1103/PhysRevD.109.026008}
    {Phys. Rev. D \textbf{109} (2024) 026008}.


\bibitem{Zhong:2022ecl}
    T.~Zhong, H.~B.~Fu and X.~G.~Wu,
    \href{https://doi.org/10.1103/PhysRevD.105.116020}
    {Phys. Rev. D \textbf{105}, 116020 (2022)}.


\bibitem{Zhong:2011rg}
    T.~Zhong, X.~G.~Wu, H.~Y.~Han, Q.~L.~Liao, H.~B.~Fu and Z.~Y.~Fang,
    \href{https://doi.org/10.1088/0253-6102/58/2/16}
    {Commun. Theor. Phys. \textbf{58}, 261-270 (2012)}.


\bibitem{Hu:2021zmy}
    D.~D.~Hu, H.~B.~Fu, T.~Zhong, L.~Zeng, W.~Cheng and X.~G.~Wu,
    \href{https://doi.org/10.1140/epjc/s10052-021-09958-0}
    {Eur. Phys. J. C \textbf{82}, 12 (2022)}.


\bibitem{Duplancic:2008ix}
    G.~Duplancic, A.~Khodjamirian, T.~Mannel, B.~Melic and N.~Offen,
    \href{https://doi.org/10.1088/1126-6708/2008/04/014}
    {JHEP \textbf{04}, 014 (2008)}.


\bibitem{Huang:1994dy}
    T.~Huang, B.~Q.~Ma and Q.~X.~Shen,
    \href{https://doi.org/10.1103/PhysRevD.49.1490}
    {Phys. Rev. D \textbf{49}, 1490-1499 (1994)}.


\bibitem{Ball:2006wn}
    P.~Ball, V.~M.~Braun and A.~Lenz,
    \href{http://doi.org/10.1088/1126-6708/2006/05/004}
    {JHEP \textbf{05}, 004 (2006)}.


\bibitem{ParticleDataGroup:2022pth}
    R.~L.~Workman \textit{et al.} [Particle Data Group],
    \href{https://doi.org/10.1093/ptep/ptac097}
    {PTEP \textbf{2022}, 083C01 (2022)}.


\bibitem{BESIII:2021bdp}
    M.~Ablikim \textit{et al.} [BESIII Collaboration],
    \href{https://doi.org/10.1103/PhysRevLett.127.171801}
    {Phys. Rev. Lett. \textbf{127}, 171801 (2021)}.


\bibitem{FlavourLatticeAveragingGroup:2019iem}
    S.~Aoki \textit{et al.} [Flavour Lattice Averaging Group],
    \href{htpss://doi.org/10.1140/epjc/s10052-019-7354-7}
    {Eur. Phys. J. C \textbf{80}, 113 (2020)}.

\bibitem{Tian:2023vbh}
    H.~J.~Tian, H.~B.~Fu, T.~Zhong, X.~Luo, D.~D.~Hu and Y.~L.~Yang,
    \href{https://doi.org/10.1103/PhysRevD.108.076003}
    {Phys. Rev. D \textbf{108}, 076003 (2023)}.

\bibitem{Hietala:2015jqa}
    J.~Hietala, D.~Cronin-Hennessy, T.~Pedlar and I.~Shipsey,
    \href{https://doi.org/10.1103/PhysRevD.92.012009}
    {Phys. Rev. D \textbf{92}, 012009 (2015)}.


\bibitem{Cheng:2017pcq}
    H.~Y.~Cheng and X.~W.~Kang,
    \href{https://doi.org/10.1140/epjc/s10052-017-5170-5}
    {Eur. Phys. J. C \textbf{77}, 587 (2017)}.


\bibitem{CLEO:2009dyb}
    J.~Yelton \textit{et al.} [CLEO],
    \href{https://doi.org/10.1103/PhysRevD.80.052007}
    {Phys. Rev. D \textbf{80}, 052007 (2009)}.


\bibitem{Na:2011mc}
    H.~Na, C.~T.~H.~Davies, E.~Follana, J.~Koponen, G.~P.~Lepage and J.~Shigemitsu,
    \href{https://doi.org/10.1103/PhysRevD.84.114505}
    {Phys. Rev. D \textbf{84}, 114505 (2011)}.



\bibitem{Lubicz:2017syv}
    V.~Lubicz \textit{et al.} [ETM],
    \href{https:// doi.org/10.1103/PhysRevD.96.054514}
    {Phys. Rev. D \textbf{96}, 054514 (2017)}.



\bibitem{HFLAV:2022esi}
    Y.~S.~Amhis \textit{et al.} [HFLAV Collaboration],
    \href{https://doi.org/10.1103/PhysRevD.107.052008}
    {Phys. Rev. D \textbf{107}, 052008 (2023)}.

\end{thebibliography}
\end{document}